# Once More About the "Forbidden" Domain Structure and the Isolated Point in $K_2Cd_{2x}Mn_{2(1-x)}(SO_4)_3$ Langbeinites

[1]Vlokh R., [1]Girnyk I., [1]Vlokh O.V., Skab [1]I., Say[1]A. and [2]Y.Uesu

[1]Institute of Physical Optics, 23 Dragomanov St., 79005 Lviv, Ukraine
[2]Department of Physics, Waseda University, 3-4-1 Okubo, Shinjuku-ku, Tokyo-160, Japan

## Abstract

The domain structure in $K_2Cd_{2x}Mn_{2(1-x)}(SO_4)_3$ (x=0.5, 0.7 and 0.9) langbeinite crystals is studied with the aid of optical polarization microscopy. It is shown that the domain walls in ferroelastic langbeinites separate enantiomorphous orientation states. These orientation states appear in the phase with the symmetry 23 in connection with the hypothetic phase transition $\bar{4}3m\ F\ 23$. In the phase with the symmetry 222, these domain walls are transformed to those separating ferroelastic domains with the opposite signs of enantiomorphism. It is revealed that one enantiomorphous domain can only transform to the other via a thick layer of the parent phase with $\bar{4}3m$ symmetry. The results for the volume thermal expansion are presented. It is shown that the isolated point at the x,T-phase diagram of $K_2Cd_{2x}Mn_{2(1-x)}(SO_4)_3$ solid solutions corresponds to the concentration x=0.6.

## Introduction

In our previous works [1-3], we have already reported on appearance of the so-called "forbidden" domain structure (according to the definition by *J.Sapriel* [4]) in the ferroelastic (FE) langbeinite crystals, which possess the phase transition (PT) with the point symmetry group change 23F222. This domain structure exists in a narrow temperature interval below $T_c$=432K in the pure $K_2Cd_2(SO_4)_3$ crystals, while in the pure $K_2Mn_2(SO_4)_3$ it is peculiar for the all temperature region of the FE phase ($T<T_c$=191K). The FE PT in $K_2Cd_2(SO_4)_3$ is of a first order close to a second-order one, whereas $K_2Mn_2(SO_4)_3$ crystals manifest a strong first-order PT. It has been also shown [5] that an "isolated point" of the second-order PT (i.e., the point of a second-order PT on a line of first-order PTs) exists at the x,T-phase diagram of

$K_2Cd_{2x}Mn_{2(1-x)}(SO_4)_3$ solid solutions at x≈0.8. This conclusion has followed from the temperature behaviours of the birefringence, thermal expansion and the dielectric permitivity near $T_c$ in the crystals with x=0.8 [5,6]. While observing the PT in $K_2Cd_{1.6}Mn_{0.4}(SO_4)_3$ using polarized microscopy, we have not found the well-defined phase boundary movement but only a wedge-like growth of FE domains into the paraelastic phase region. Nevertheless, we have not observed the appearance of FE phase in the whole sample at a unique temperature, as it should have been at a second-order PT. Such the behaviour may be caused by the fact that the composition with x=0.8 does not exactly corresponds to the isolated point. In order for checking the last assumption, we have started studies of the compositions close to x=0.8. This is why, one of the goals of the present paper is to investigate the phase boundary and the domain structure of $K_2Cd_{2x}Mn_{2(1-x)}(SO_4)_3$ (x=0.5, 0.7 and 0.9) crystals and the volume thermal expansion in all of the said solid solutions. Another problems of the "forbidden" FE domains in langbeinites, which have not been yet solved, are as follows: why the domain walls are so thick (some tens of micrometers) and, moreover, optically isotropic?

### Experimental

We studied domain structure of $K_2Cd_{2x}Mn_{2(1-x)}(SO_4)_3$ crystals at the PT with the help of optical polarization microscope, using a cooling cell that permitted controlling temperature down to the liquid-nitrogen temperatures with the precision close to 0.1K. In some cases, we used a compensator with a small optical retardation for distinguishing the areas weakly differed by the birefringence. Photographing was performed with a standard photo camera. $K_2Cd_{2x}Mn_{2(1-x)}(SO_4)_3$ single crystals were grown with the Bridgman technique. The crystalline plates of <001> orientation with the thickness of 0.3mm were cut off the bulk samples with a diamond wire and polished with a diamond paste. Studies of the volume thermal expansion were carried out with the aid of a capacity dilatometer.

### Results and discussion

It is well known that, according to the *Landau* theory of PT (see, e.g., [7]), the volume expansion in the FE phase due to a second order FE PT is proportional to the change of temperature,

$$\frac{\Delta V}{V} \propto e_s \propto \Theta^2 \propto (T - T_c), \qquad (1)$$

where $e_s$ is the spontaneous deformation and $\Theta$ the order parameter. Furthermore, the temperature change in the volume expansion should be discontinuous in the vicinity of the second-order PT point $T_c$. Thus, the dependence of thermal expansion coefficient $\beta$ upon temperature should exhibit a break or a jump near $T_c$ rather than a peak-like anomaly. As seen from Figure 1, the anomalous part of the volume expansion coefficient approaches zero somewhere in the range between x=0.5 and x=0.7 but not at x=0.8. After fitting the concentration dependence of anomalous part of the volume thermal expansion by the third-power polynomial, it follows that the curve manifests a minimum at x=0.6. As a matter of fact, the anomalous part of the volume expansion coefficient is not exactly equal to zero at x=0.6. It has a small value $\Delta\beta=0.7\times10^{-4}$ at that point. Thus, one can conclude that the coordinates of the isolated point at the x,T-phase diagram (or, at least, the point which is very close to the isolated one) are equal to (0.6; 213K).

In the cooling regime, appearance of the FE phase in $K_2Cd_{1.8}Mn_{0.2}(SO_4)_3$ crystals at $T_c$=363K is accompanied with a movement of phase boundaries from the side regions of samples towards the centre (see Figure 2). The appearance of the phase boundary testifies unambiguously the fact that the PT in $K_2Cd_{1.8}Mn_{0.2}(SO_4)_3$ is of the first order. The following cooling run brings to transformation of central paraelastic layer in the sample into two FE domain states, which are separated by a thick domain wall. The region of this domain wall remains optical isotropic (cubic) down to the room temperature. Moreover, the phase boundaries that have appeared at $T_c$ are transformed to the domain walls at cooling. The latter are inclined to (001) plane. Such the multidomain structure, which includes a few separated domains and the domain walls, exists down to the room temperature. It is interesting to notice that the description of the domain structure observed in $K_2Cd_{1.8}Mn_{0.2}(SO_4)_3$ crystals, based on our previous approach [1] alone, would be impossible.

The FE phase in $K_2Cd_{1.4}Mn_{0.6}(SO_4)_3$ crystals appears at $T$=285K in the cooling regime (see Figure 3). It is worthwhile that the appearance of FE phase in some spatial region of sample (see the right part of Figure 3) is accompanied by nucleation of a spot-like structure with the ill-defined boundaries. Such the behaviour is typical for second-order PTs. At the same time, the movement of the phase boundary is observed in the left part of sample. Hence, the crystals with x=0.7 are very close to the conditions that correspond to a second-order PT. Let us notice also that a well-defined phase boundary is observed in $K_2CdMn(SO_4)_3$ crystals (Fig. 3).

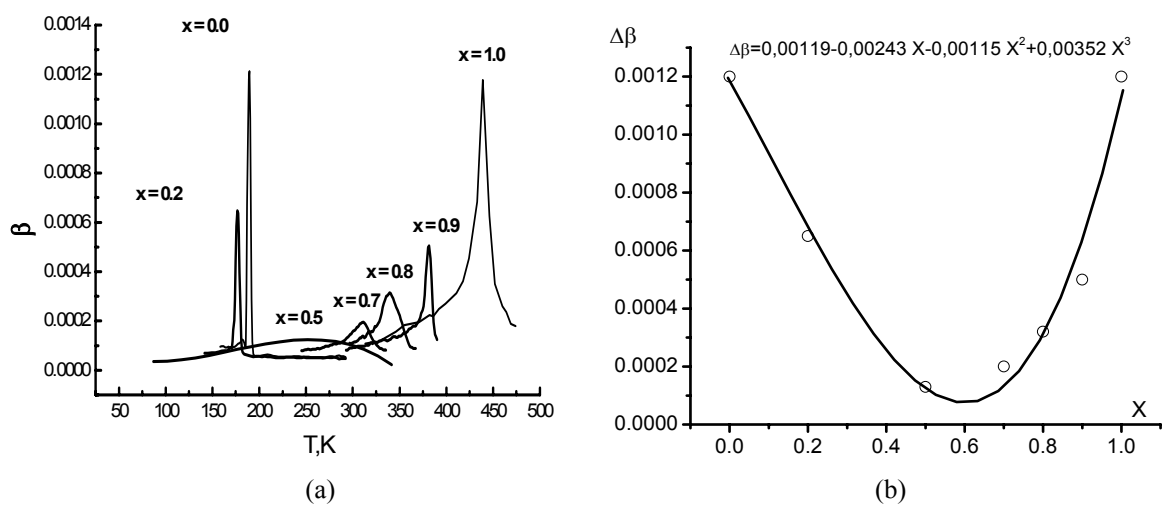

**Fig. 1.** Temperature dependences of the volume thermal expansion in $K_2Cd_{2x}Mn_{2(1-x)}(SO_4)_3$ crystals (a) and concentration dependence of the magnitude of peak-like anomalous part of the volume thermal expansion (b).

In our previous papers (see, e.g., [1]), we have described a orientation of the "forbidden" domain walls in langbeinites, basing on the concept of parent phase with the point group of symmetry $\bar{4}3m$. The sequence of high-order ferroic and FE PTs in this case is as follows: $\bar{4}3m \, F \, 23 \, F \, 222$. This assumption suggests that the phase 23 can possess two orientation states (one "right" and the other "left"), while the FE phase 222 – six FE orientation states (three of them "right" and three others "left"). Let us remind that a rank-two polar tensor of spontaneous deformation is not sensitive to the change of sign of the coordinate system. Therefore, the condition of elastic compatibility ($\det \Delta e_s = 0$) cannot be satisfied even under increasing the number of orientation states up to six. Moreover, the next assumption made in [1] ($Sp(e_s) \approx 0$) is also not successful enough, because there are no FE langbeinites, for which one of diagonal components of spontaneous deformation tensor is exactly equal to zero. Nevertheless, the FE domain walls exist in many FE langbeinites, e.g., in $Tl_2Cd_2(SO_4)_3$ [8], $K_2Co_2(SO_4)_3$ [9] and all of $K_2Cd_{2x}Mn_{2(1-x)}(SO_4)_3$ crystals. In all the known cases of 23F222 PTs, the domain walls have the same {110} orientation. In spite of this, the assumption about existence of the parent phase with the symmetry $\bar{4}3m$ is indeed correct, since we have observed enantiomorphous regions in the phase with the symmetry 23 in $K_2Cd_{0.4}Mn_{1.6}(SO_4)_3$ crystals [10]. It follows from the mentioned experimental facts that the domain walls appear between the two enantiomorphous domains already in the phase with the symmetry 23. The domain wall may have only {110} orientation (i.e., the orientation of the mirror planes lost at the hypothetic PT $\bar{4}3m \, F \, 23$ – see Fig. 4).

Then, the domain walls separating the orientation states with, at least, the same sign of spontaneous deformation but opposite signs of enantiomorphism would be permissible at the PT to FE phase. The orientation of these domain walls should remain {110} at the 23F222 PT.

The two questions formulated above still arise in frame of this approach: why the domain walls are comparatively thick (tens of micrometers) and why they manifest a cubic symmetry?

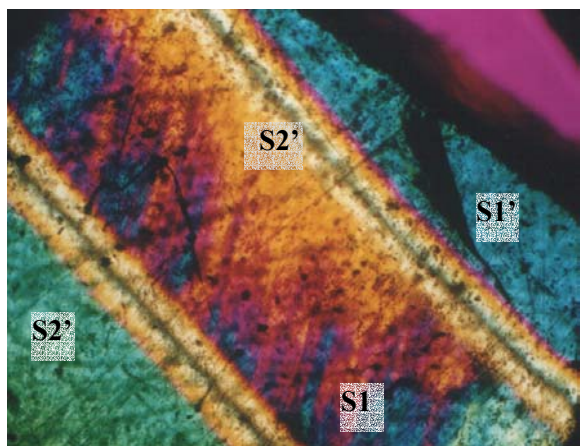

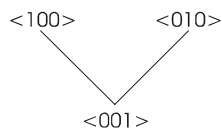

**Fig. 2.** Domain structure in $K_2Cd_{1.8}Mn_{0.2}(SO_4)_3$ crystals at 343K.

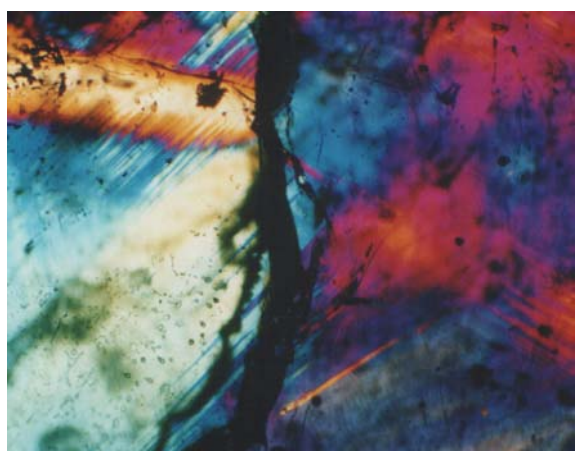

**Fig. 3.** Nucleus of FE phase (blue spots) in the paraelastic matrix (violet spots) of $K_2Cd_{1.4}Mn_{0.6}(SO_4)_3$ crystals and the phase boundary in $K_2Cd_{0.5}Mn_{0.5}(SO_4)_3$ at $T_c$.

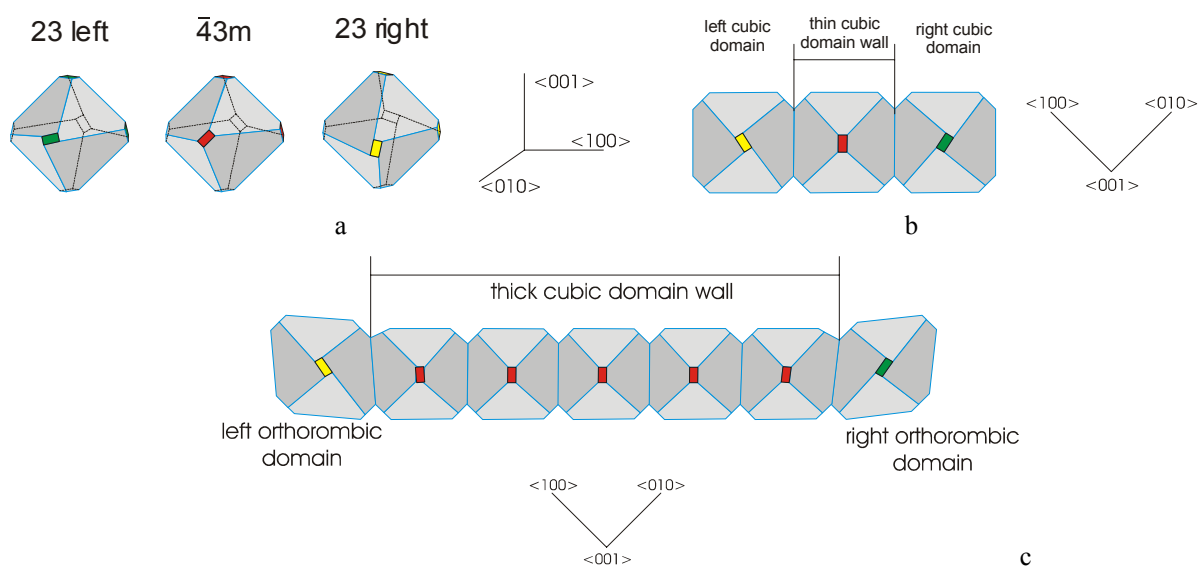

**Fig. 4.** Habit of the unit cells in the cubic phases (a), "left" and "right" domains in the phase 23 (b), and "left" and "right" domains in the FE phase 222 (c).

In this respect, one can remind of ferromagnetic domain walls, which are also "thick" because of a gradual change in the orientation of spontaneous magnetization pseudo-vector inside the wall (see, e.g., [11]). Inside a ferromagnetic domain wall, the magnetization pseudo-vector changes its orientation but never becomes zero, while the spontaneous polarization vector and the spontaneous deformation tensor inside a ferroelectric or FE domain wall achieve a zero value in the geometrical centre of the wall and change their sign there. The domain walls in ferromagnetics and high-order ferroics, which are characterized by appearance of optical activity in a low-temperature phase, possess some common properties – they separate enantiomorphous domains. The difference is that, in case of ferromagnetics, the domains differ by the sign of pseudo-vector, while in our case by the sign of a second-rank pseudo-tensor. It is worth noticing that single-domain states in small-sized samples of ferromagnetics are energetically preferable. The small samples of $K_2Cd_2(SO_4)_3$ crystals are also single domain [2]. Because of impossibility of compensation of spontaneous magnetization with external magnetic field (it would be impossible to stop a charge movement, and, in the presence of ordering external field, the magnetization cannot be compensated by means of disordering magnetic moments), the switching process may take place only through a gradual rotation of magnetization vector, beginning from the orientation peculiar for one domain and ending by that of the opposite domain. The magnetization existing at each point of ferromagnetic domain wall would lead to appearance of mechanical deformations, due to magnetostriction effect. In order to satisfy elastic compatibility in the layers with gradually changed deformations, the domain wall becomes "thick" enough.

In ferroelectrics, one can compensate the spontaneous polarization while applying a biasing electric field in the process of domain switching. The spontaneous deformation in FEs should gradually decrease to zero inside the wall and then reappear, with the opposite sign in the neighbouring domain. Let us now analyse how to pass from a "left" domain to a "right" one in the 23-phase, under the condition that the crystal should remain continuous. Even in the 23-phase, in order to ensure elastic compatibility of enantiomorphous domains in the course of transition from "left" to "right" domain, it is necessary to pass gradually through a layer of parent phase with the symmetry $\bar{4}3m$ (it correspond to the zero value of order parameter inside the wall) (see Figure 4b). Thus, only the layer with the symmetry $\bar{4}3m$ can play a role of domain wall in the 23-phase, which appears owing to $\bar{4}3m\ F\ 23$ PT. Moreover, only domain walls with this cubic symmetry should exist in the FE phase, which alone can separate the neighbouring domains with the same spontaneous deformation tensor but the opposite signs of enantiomorphism. This is the reason why we have always observed isotropic layers as the domain walls in FE langbeinites. Furthermore, the domain walls in the FE phase become thick enough, because the lattice parameters of the cubic ($\bar{4}3m$) and FE (222) phases differ larger than those of the $\bar{4}3m$ and 23 phases, as a result of spontaneous deformations (see Figure 4c).

Now one can easily explain the domain structure observed in Figure 2. If we suppose that the crystal should "return" to the point symmetry group $\bar{4}3m$ inside the domain wall, then, from the viewpoint of energy preferring, it does not matter which particular FE domain would appear on the opposite side of the wall. There is only one limitation – these domains should be enantiomorphous. Thus, six different domain walls can exist in the phase 23, with {110} orientation. The number and orientations of the corresponding walls would coincide with those of the mirror planes lost in the course of $\bar{4}3m\ F\ 23$ PT. In the FE phase, these domain walls may separate any FE domains, which are enantiomorphous. Denoting the "right" FE domains as S1, S2 and S3 and the "left" as S1′, S2′ and S3′, one can arrive at the conclusion that there are three different orientation states in Figure 2. In the central part, the thick domain wall with (110) orientation separates two domains with different spontaneous deformation, birefringence, colour (yellow and blue) and signs of enantiomorphism (let us indicate them as S1 (blue) and S2′ (yellow)). The walls inclined to (001) plane have (101) orientation. They separate S2′ and S1 domains, as well as S1 and S1′ domains. The S1 and S1′ domains are of the same blue colour. It means that their deformations are the same and the enantiomorphism signs opposite. In the right bottom part of Figure 2, one can see a small region of residual domain walls between the identical S1 and S1 domains. Probably, the domain wall that separates these same domains should disappear at further cooling.

## Conclusions

We have shown in the present paper that the domain walls in FE langbeinites are just those that separate enantiomorphous orientation states appearing in the phase characterized with the symmetry 23. They are associated with the hypothetic $\bar{4}3m\,F\,23$ PT. We have also explained why these domain walls are relatively thick and optically isotropic. On the basis of the volume thermal expansion studies it has been shown that the PT in $K_2Cd_{2x}Mn_{2(1-x)}(SO_4)_3$ crystals is close to the second-order at x=0.6. The observed domain structure of $K_2Cd_{1.8}Mn_{0.2}(SO_4)_3$ crystals is in a good agreement with the explanations suggested above.

## Acknowledgement

The authors are grateful to the Ministry of Education and Science of Ukraine (the Project N0103U000700) for the financial support of this work.

## References

1. Vlokh R., Uesu Y., Yamada Y., Skab I., Vlokh O.V. J. Phys. Soc. Jap. (Lett.) **67** (1998) 3335.
2. Vlokh R., Kabelka H., Warhanek H., Skab I., Vlokh O.V. Phys. Stat. Sol.(a) **168** (1998) 397.
3. Vlokh R., Skab I., Vlokh O., Uesu Y. Ukr. J. Phys. Opt. **2** (2001) 148.
4. Sapriel J. Phys. Rev. B **12** (1975) 5128.
5. Vlokh R., Vlokh O., Kityk A., Skab I., Girnyk I., Czapla Z., Dacko S., Kosturek B. Ferroelectrics **237** (2000) 481.
6. Vlokh R., Czapla Z., Kosturek B., Skab I., Vlokh O.V., Girnyk I. Ferroelectrics **219** (1998) 243.
7. Smolenskiy G.A. et. al. Physics of Ferroelectric Phenomena. Leningrad, Nauka, (1985) 396p. (in Russian).
8. Vlokh R., Skab I., Girnyk I., Czapla Z., Dacko S., Kosturek B. Ukr. J. Phys. Opt. **1** (2000) 28.
9. Brezina B., Rivera J.-P., Schmid H. Ferroelectrics **55** (1984) 177.
10. Vlokh R., Skab I., Vlokh O., Yamada Y. Ferroelectrics **242** (2000) 47.
11. Vonsovsky S.V. Magnetism, Moscow, Nauka (1971) 1032p. (in Russian)